\documentclass[preprints,article,accept,moreauthors,pdftex,10pt,a4paper]{Definitions/mdpi}

\usepackage{framed}
\usepackage[strict]{changepage}    

\definecolor{mypink1}{rgb}{0.858, 0.188, 0.478}
\definecolor{mypink2}{RGB}{219, 48, 122}
\definecolor{mypink3}{cmyk}{0, 0.7808, 0.4429, 0.1412}
\definecolor{mygray}{gray}{0.8}

\definecolor{formalshade}{rgb}{0.95,0.95,1}    

\newenvironment{formal}{%
  \MakeFramed{\advance\hsize-\width\FrameRestore}%
  \noindent\hspace{-4.55pt}
  \begin{adjustwidth}{}{7pt}%
  \vspace{2pt}\vspace{2pt}%
}
{%
  \vspace{2pt}\end{adjustwidth}\endMakeFramed%
}

\usepackage{physics}

\usetikzlibrary{shapes.geometric,calc}

\firstpage{1} 
\pubvolume{xx}
\issuenum{1}
\articlenumber{1}
\pubyear{2018}
\copyrightyear{2018}
\externaleditor{Academic Editor: name}
\history{Received: date; Accepted: date; Published: date}

\Title{From the digital data revolution toward a digital society: Pervasiveness of Artificial Intelligence}

\Author{Frank Emmert-Streib $^{1,2,*}$\orcidA{} }

\AuthorNames{Frank Emmert-Streib}

\address{%
$^{1}$ \quad Predictive Society and Data Analytics Lab, Faculty of Information Technology and Communication Sciences, Tampere University, Finland\\
$^{2}$ \quad Institute of Biosciences and Medical Technology, Tampere, Finland\\
}

\corres{Correspondence: v@bio-complexity.com; Tel.: +358-50-301-5353}

\abstract{Technological progress has led to powerful computers and communication technologies that penetrate nowadays all areas of science, industry and our private lives. As a consequence, all these areas are generating digital traces of data amounting to big data resources. This opens unprecedented opportunities but also challenges toward the analysis, management, interpretation and utilization of these data. Fortunately, recent breakthroughs in deep learning algorithms complement nowadays machine learning and statistics methods for an efficient analysis of such data. Furthermore, advances in text mining and natural language processing, e.g., word-embedding methods, enable the processing of large amounts of text data from diverse sources such as governmental reports, blog entries in social media or clinical health records of patients. In this paper, we present a perspective on the role of artificial intelligence in these developments and discuss potential problems we are facing in a digital society. 
}


\begin{document}


\section{Introduction}

In the last few decades, technological progress has changed nearly all areas of science \cite{chang2014understanding,chen2012business,helbing2015thinking}. This comprises many fields, including biology, computer science, economy, engineering, humanities, journalism, politics, public health, management, medicine, social sciences, sports and even arts. While the generation of data has long been a privilege of basic research, the computerization of society and the establishment of the Internet have enabled the availability and the distribution of information and data on almost all aspects of our daily lives. As a consequence, a quantitative analysis of such digital data can be conducted by means of artificial intelligence (AI) and machine learning with results that might have a profound effect on all levels of society \cite{Maayan_2014,olshannikova2017conceptualizing,kitchin2014data}.

A field that was among the first transitioning into a technology-driven area was biology \cite{schena1995quantitative,marx_2013}. Importantly, the Human Genome Project \cite{quackenbush_2011} helped in enhancing molecular high-throughput measurements, e.g, next-generation sequencing (NGS) technologies \cite{shendure_2008}, which allows the interrogation of all molecular levels, including mRNAs, proteins and DNA sequences \cite{em_be6,em_jSep0109}. In recent years, this technology has also infiltrated the biomedical and clinical sciences which allowed a quantification of those fields as well. Further areas that have been significantly transformed by the digital revolution are economy and business. Importantly, most of the trading on the stock markets worldwide is nowadays conducted electronically, i.e., orders can be placed online and are directly sent to a broker circumventing traditional floor trading. Lastly, also the social sciences have been heavily influenced by big data \cite{olshannikova2017conceptualizing,shah2015big}. For instance, various kinds of social media platforms, e.g, facebook, twitter or linkedin, provide a sort of virtual laboratories for conducting social and psychological experiments leading to novel insights of human behavior \cite{conte2012manifesto,em_jJul012018,matz2017psychological}. 

Considering the fact that technologies like the Internet, NGS or the iPhone have only been available since 1991, 2004 and 2007 respectively it seems clear that within the next thirty years the pace of new inventions will likely further increase. Hence, novel technologies building upon existing once will further transform science, industry and our private lives in profound and potentially hard to foreseeable ways. In this paper, we accept this challenge and take a look ahead by discussing some of the potential changes for these fields, beneficially or detrimentally. Our final outlook will address {\it digital society} and we will argue that this state is obtained as an asymptotic limiting state of digital economy characterized by the pervasiveness of artificial intelligence.

This paper is organized as follows. In the next sections, we discuss several fields that have been significantly reshaped by the digital revolution. We discuss opportunities for the method development in artificial intelligence and potential domain-specific challenges. Then we discuss general instances of artificial intelligence approaches and learning paradigms that might be especially beneficial to all fields effected by the digitalization. Thereafter, we discuss important aspects of future challenges that are of crucial importance for the development of the respective fields. For this discussion, and the previous presentation, we assume an AI-centered perspective. Aside from mostly positive effects of a digitalization, serious problems thereof are addressed, e.g., about data privacy and fundamental issues of artificial intelligence governance. As a key problem, the asymptotic state of digital economy, we call digital society, is discussed. The paper finishes with concluding remarks.

\section{Digital medicine and digital health}

As already mentioned, biology experienced a transition toward a technology-driven area in the 1990s. This was accomplished by introducing the DNA microarray technology allowing the measurement of genome-scale information of the concentration of messenger ribonucleic acids (mRNAs). Further technologies that followed were SELDI (Surface-enhanced laser desorption/ionization), protein-chips and various forms of NGS assays (next-generation sequencing), e.g., DNA or RNA sequencing or DNA methylation \cite{petricoin2004seldi,marzese2015emerging,wang_2009}. Importantly, many of these technologies also propagated to medical, clinical and public health studies which made also these fields essentially data-driven as a consequence of such technologies. 

There are many subfields of the above subjects that utilize modern information and communication technologies in biology, medicine and public health. However, the terms digital medicine or digital health are commonly used to indicate the general integration of such digital technologies, e.g., with smartphone or sensor technologies, with advanced analysis methods to enhance the subject related goals \cite{chen2019characteristics,steinhubl2018digital,kostkova2015grand}. Interestingly, in \cite{fogel2018artificial} it is noted that "Despite a flurry of recent discussion about the role and meaning of AI in medicine, in 2017 nearly $100\%$ of U.S. healthcare will be delivered with $0\%$ AI involvement." This statement underlines the difficult road ahead for translating results from basic research to the application in hospitals or healthcare systems but shows also the potential for methods from AI.

{\bf Challenges and obstacles:} A necessity for AI to make beneficial contributions to medicine and health, but also to other fields discussed below, is the availability of (large amounts of) data. However, present genomics technologies, clinical and pathological imaging technologies, biosensors, and the internet of things (IoT) devices are essentially capable of fueling AI methods with sufficient data. There is just the requirement to gather patient-specific data over a longer period of time to establish data repositories similar to the ImageNet database for images. Then personalized or patient-tailored methods can be developed and benchmarked to enhance the current state-of-the-art in computational diagnostics and evidence based medicine. 

 Promising pilot studies exist that demonstrate the utility of AI methods, especially deep learning, for digital health. For instance, such studies were conducted for diabetic retinopathy \cite{gulshan2016development}, skin cancer \cite{esteva2017dermatologist} and medication adherence \cite{labovitz2017using}, to name just a few. Interestingly, most of such studies are mainly based on image analysis. This is another indicator of the early stage of digital medicine because medicine and health offer many more data types, as mentioned above, beyond imaging data.

There are three major concerns frequently raised against AI in medicine and health. The first is the fear that jobs will be lost due to the introduction of automatic analytics systems, the second criticizes the potential disruption of the personal doctor-patient relationship for similar reasons and the third issue relates to the lack of explainability of general AI methods \cite{fogel2018artificial,emmert2020explainable}. The latter point means that usually AI models can be considered as {\it black-box} prediction models that are capable of achieving high prediction performance but lack intuitive explanations that describe, e.g., in standard medical terms, how the performance was actually obtained. 

It is important to highlight that all three concerns do not relate to methodological issues of AI itself but to job safety, trust and communication. This means in order to pave the way for AI in medicine and health there is also educational work of the public and control bodies necessary to overcome negative and possibly ill-informed sentiments. Furthermore, it is important to mention that digital health and digital medicine require a multidisciplinary approach for their successful deployment \cite{kostkova2015grand}. Given the experience of similar but potentially smaller-scale endeavors from bioinformatics or systems biology there is already demonstrated success one can build on in forming the cross- and interdisciplinary teams needs.

Finally, for all the above approaches there are privacy and ethical issues that need to be dealt with properly \cite{vayena2018digital,milosevic2019ethics}. On one hand, this needs to ensure that a patient is in control of its own data but also that sufficient data are collected and available for the development of data-driven AI methods. This is certainly a none-trivial balancing act to fulfill all needs. A potential circumvention of this could be the anonymization of patient data in a way that the data are modified in a way that individuals are no longer identifiable yet the modified data are not effecting prediction results from AI methods. For some promising pilot studies, see \cite{lee2017utility,el2015anonymising}.

\section{Digital economy and business}

The digitalization of our world is not limited to medicine and health but effects also the way we conduct business and our entire economy. There have been many attempts to define 'digital economy' and a nice review of a large number of such definitions can be found in \cite{bukht2017defining}. Overall, the common agreement about the nature of digital economy is succinctly summarized by "...an economy based on digital technologies (sometimes called the internet economy)" provided by the 
{\it Expert Group on Taxation of the Digital Economy} of the European Commission. We would like to note that sometimes digital economy is all called {\it web economy} or {\it new economy}.  A similar definition for digital business has been provided by \cite{bharadwaj2013digital} as "organizational strategy formulated and executed by leveraging digital resources to create differential value". Both definitions are rather general but the diversity of these fields requires such a wide characterization to encompass all relevant aspects thereof.

Specific main sectors that are included in the above definitions are:
\begin{itemize}
\item e-Business
\item e-Commerce
\item Industry 4.0
\item Sharing economy
\item Crowdsourcing
\end{itemize}

Here electronic business (e-business) "is any process that a business organization conducts over computer-mediated networks" and electronic commerce (e-commerce) "is the value of goods and services sold over computer-mediated networks" \cite{mesenbourg2001measuring}. Industry 4.0 (or smart factory) stands for the fourth industrial revolution which is transforming traditional manufacturing and industrial processes into a technology mediated field \cite{lu2017industry,xu2018industry}. This includes machine-to-machine communication (M2M), Internet of Things (IoT) and cyber-physical systems (CPS). According to Kagermann et al. \cite{kagermann2013recommendations}, Industry 4.0 is "a new level of value chain organization and management across the lifecycle of products." That means not only the production and manufacturing is effected but also decision making across all relevant levels including the management. Furthermore, Industry 4.0 includes not only customization but also a personalization of products \cite{wang2017industry}.

\begin{table}[t!]
  \small
  \begin{center}
    \caption[]{Key technologies for Industry 4.0. For a discussion of succinct differences and commonalities between these technologies see \cite{chen2012machine,wan2013machine}. }
    \label{tab.sum}
    \begin{tabular}{ l | p{6cm} c }
      \hline
      \textbf{Technology} & \textbf{Definition} & \textbf{Reference}\\ 
      \hline
      \hline
     Machine-to-machine communication (M2M) & "Machine-to-Machine  (M2M)  paradigm  enables  ma-chines  (sensors,  actuators,  robots,  and  smart  meter  readers)  tocommunicate with each other with little or no human intervention.M2M is a key enabling technology for the cyber-physical systems(CPSs)." & \cite{stojmenovic2014machine}   \\
  Wireless sensor networks (WSN)   & "WSN is designed particularly for delivering sensor-related data." & \cite{wu2011wireless} \\
     Internet of Things (IoT) &   "An open and comprehensive network of intelligent objects that have the capacity to auto-organize, share information, data and resources, reacting and acting in face of situations and changes in the environment." & \cite{madakam2015internet} \\
      Cyber-physical systems (CPS) &    "CPS are systems of collaborating computational entities which are in intensive connection with the surrounding physical world and its on-going processes, providing and using, at the same time, data-accessing and data-processing services available on the internet."        & \cite{Monostori2018}     \\
      \hline
    \end{tabular}
  \end{center}
\end{table}

Another important part of digital economy is sharing economy (SE). Sometimes SE is also called access economy, peer-to-peer (P2P) economy or collaborative economy \cite{cheng2016sharing,botsman2010s}. Also SE is a wide term that has been defined as: "the sharing economy is an IT-facilitated peer-to-peer model for commercial or non-commercial sharing of underutilized goods and service capacity through an intermediary without a transfer of ownership" \cite{schlagwein2020consolidated}. Hence, its underlying idea is to directly connect individual consumers and individual providers of goods or services facilitated by a community-based on-line platform. Examples of such business models are:
\begin{itemize}
\item freelancing platforms (labor market consisting of short-term contracts)
\item coworking platforms (individuals working independently or collaboratively in shared office space)
\item P2P lending platforms
\item fashion platforms
\end{itemize}

Finally, crowdsourcing (CS) shares some similarity to sharing economy and it is defined as: 
"Crowdsourcing  is  the  IT-mediated  engagement  of  crowds  for  the  purposes  of  problem-solving,  task completion,  idea  generation  and  production" \cite{taeihagh2017crowdsourcing}. Important examples of CS are information sharing systems, e.g., Wikipedia or del.icio.us, voting systems, e.g., Amazon's Mechanical Turk (MTurk) and gamification, e.g., reCAPTCHA (image recognition) or Foldit (protein folding).

In order to show the economic importance of the above fields on economy itself several studies have been conducted. For instance, sharing economy is in $2020$ valued at US$\$15$ billion globally with a potential to raise its global market value to US$\$335$ billion by $2025$ \cite{lim2020sharing}. Measuring the value of digital economy by a study of the United Nations estimated that digital economy contributes $4.5\%$ to $15.5\%$ of the world GDP \cite{unctad2019digital}. These numbers show the enormous impact of digital economy and its comprising subfields on our world economy and its potential to further increase.

{\bf Challenges and obstacles:} For AI systems there is a large number of directions open for contribution. In general, artificial intelligence should be central for any data-driven approach in digital economy including Industry 4.0. For instance, AI can make valuable contributions to predictive maintenance (PdM) \cite{grall2002continuous,lee2019predictive}. PdM is dealing with maintenance issues of production devices or general machines and helps in reducing down times or operational costs. By utilizing sensor information of either production or operation lines AI-based prediction models can be trained and used for optimizing maintenance schedules. Furthermore, AI should be helpful for any kind of IoT or CPS application because such technologies were designed for the gathering of data but not for their analysis. Finally, AI can help to further improve robotics and automation for manufacturing, production or service applications. Currently, deep reinforcement learning is showing promising results for such a novel AI approach in this context \cite{mnih2015human,gu2017deep}.

On a more fundamental point, it is important to note that the application of AI requires also an adjustment regarding general data analysis principles. An early standard of this has been called CRISP-DM (cross-industry standard process for data mining) \cite{shearer2000crisp} emphasizing the feedback between consecutive analysis steps. Recently, this has been extended considering also industry-specific needs and domain-specific knowledge \cite{em_tripathi2020ensuring}.

There are three major concerns frequently raised against AI methods in business and economy. The first is the fear that some jobs will be lost due to the introduction of automatic analytics systems \cite{brynjolfsson2014second}, the second issue relates to the lack of explainability of general AI methods and the third is concerned with the increasing gap between developed and developing countries and the general change of wealth distribution \cite{bughin2018notes}. Interestingly, the first two points are essentially the same as for digital medicine and health systems; discussed above. The latter issue is addressed by artificial intelligence governance

\section{Pervasiveness of artificial intelligence }

All of the above problems can only be studied by using methods to learn from data \cite{abu2012learning}. It is important to realize that the methods to study such problems are not always the same but they need to be adopted or newly developed to fit a given data set optimally. Hence, there is a constant need to further enhance and extend the existing pool of machine learning and artificial intelligence methods because the technology underlying the problems which enables the generation of data is constantly changing.

For reasons of clarity, we would like to emphasize that there are many scientific areas dealing with the development of novel methods for the analysis of data. For instance, machine learning, statistics, pattern recognition or artificial intelligence are all different fields with their own history and preferences for methodological approaches and conceptual frameworks \cite{theodoridis_2003,hastie_2009,em_emmert2020clarification}. However, in this paper, we simplify the discussion by summarizing these fields by the term {\it artificial intelligence} because, especially, in industry this term has become the commonly accepted standard when speaking about data analysis methods and approaches. Nevertheless, one should be aware that on the academic side this is seen differently.

Due to the diversity of fields generating general data, one can expect that the methods for their analysis are similarly diverse. Currently, methods for image and audio processing \cite{krizhevsky2012imagenet,lee2009unsupervised} seem to be much more developed than methods for other types of data, e.g., text data, genomics data or sensor data. Hence, such data types offer a great potential for improvements. For instance, for text data a fundamental problem is a conversion of textual information into numbers in a way that conventional AI methods can process such data.  Recently, word-embedding methods made great progress, above all word2vec or BERT  \cite{w2v2,bert,em_perera2020named}. However, there is still room for improvement especially for mapping onto larger units, e.g., paragraphs or documents \cite{lau2016empirical}.

Another area of great potential is information fusion \cite{mahler2007statistical}. The general idea is to combine data from multi-sensors, multi-sources or multi-processes in a way that the resulting data set contains more information than its separate, individual sources. This problem becomes apparently more difficult the more different the individual data sources or sensors are, especially, if these correspond to different data types, e.g., image data and text data. 

Also transfer learning \cite{pan2009survey} should be mentioned as a field of great potential. Transfer learning means that one starts training a model for one task and then switches the data for another one. For instance, in image processing a model that has been trained for classifying non-medical images could be transferred to learning to discriminate tumors in medical images \cite{huynh2016digital}. This example demonstrates also that transfer learning is particularly useful when one has only a very limited amount of data for a certain task (for instance from medical images) but a much larger data set for a similar task (for non-medical images).

Finally, a field that should be established is {\em statistical artificial intelligence} (SAI). SAI would extend the ideas of statistics to artificial intelligence, e.g., by investigating the influence of the sample size on the resulting prediction performance. This is important because a method in isolation is neither good nor bad but only in combination with data of certain characteristics a method can be evaluated. For instance, in \cite{em_emmert2020introductory} a deep learning classifier (a Long Short-Term Memory (LSTM) model) has been studied for classifying handwritten digital characters provided by the EMNIST (Extended MNIST) data \cite{cohen2017emnist}. As a result, it has been found that by using over $200,000$ training samples the classification error is far below $5\%$, while for $5,000$ training samples the error increases to over $30\%$. Considering that the underlying deep learning classifier was the same for both approaches, this demonstrates the importance of quantifying the influence of the sample size on the resulting performance. Formally, such a characterization of a model is obtained by so called {\it learning curves} \cite{em_jemmert2019evaluation}. In general, AI methods in digital business and digital health appear to be studied in a less stringent way as compared to, e.g., methods from biostatistics. This is understandable given the fact that the latter methods find regular application in medical and clinical patient data. Nevertheless, also in those fields a steady control is required for ensuring quality standards \cite{em_emmert2019ensuring} because there are examples violating such standards can jeopardize the lives of patients \cite{baggerly_2009}.

\section{Discussion}

The above presentation discussed the individual fields separately and focused mainly on their core components. Despite this, one could already recognize that there are many commonalities among the different fields and approaches. For this reason, in the following we focus on common aspects shared by these fields presented in an AI-centric way.

\subsection{Smart cities and smart government}

One may wonder if there are other fields or areas beyond medicine, health, business and economy that could benefit from a digitalization and utilization of AI in a similar way as, e.g., digital economy? In fact, there are already some developments in this direction. For instance, smart city and smart government are attempts to improve the organization of cities or governments respectively. In order accomplish this, smart cities utilize many sensors throughout the city, e.g., via IoT technology, to improve traffic management, road safety or energy efficiency \cite{djahel2014communications,barba2012smart,ejaz2017efficient} whereas smart government utilizes mainly automation for administrative tasks (e-government) and for data-driven decision making \cite{kankanhalli2019iot}. However, the developmental state of these areas is considerably behind other fields, e.g., digital health. One reason for this may be the fact that, traditionally, neither cities nor governments are based on electronic communication technologies. Hence, there is, first, a need to introduce information processing and computing technologies for generating and gathering data and then AI-based solutions can be designed for particular tasks. 

For the near future, for smart cities it is expected that more sensors are needed throughout the cities for gathering information about traffic, environmental conditions and human behavior. For smart government, text mining based AI approaches seem very promising because, essentially, all administrative tasks involve text data. Similar to digital economy, also smart cities and smart government will gradually develop toward higher states of digital cities and digital government. However, for the latter it remains to be seen how much involvement of AI is desirable or acceptable because, ultimately, even political decisions could be made based on such methods. 

It is certainly intriguing to think about such possibilities even if only applied retrospectively. An example demonstrating the problems with such an approach is the Brexit. Assuming we would have an AI system that would allow us to answer the following political question:
\begin{itemize}
 \item Question: Should the UK leave the European Union? 
 \end{itemize}
 Phrased like this it is a binary classification task which could be solved by a supervised learning method because the answer is either to leave or to stay in the European Union. A problem is that contrary to general artificial intelligence or machine learning approaches, there are no samples of similar 'events' available one could use for the training of this supervised learning task. Leaving technical difficulties aside, assuming we would have access to an AI system that could provide a faithful answer to this question, what would this entail? Would this answer be convincing to the people who's opinion was the opposite? Given the fact that a general classification result does not come with an explanation, such an answer could be misunderstood or even bewildering for the public in large. 
 
 This example demonstrates potential limitations even of error-free AI systems capable of making the correct political decisions. Hence, in a such a context, AI by itself could not provide the final solution but needs to be complemented with additional features not unlike to what is currently discussed for explainable AI (XAI) \cite{xu2019explainable,adadi2018peeking,emmert2020explainable,holzinger2018current}. Overall, this example shows that AI-driven decision making on higher levels, e.g., on a management or governmental stage, possess new challenges that need to be addressed.

\subsection{Human-machine interaction}

A related topic, but coming from a different perspective, is human-machine interaction (HMI) \cite{gorecky2014human}. In HMI, also called human-computer interaction (HCI) \cite{card2018psychology}, one assumes that a machine or computer is not capable of performing the complete task by its own. Instead, some form of human-involvement is needed for a succesful implementation and execution of the task. That means there is an interface between people and machines or computers. 

Prominent application examples are doctor-in-the-loop for supporting medical decision making by health practitioners \cite{holzinger2016interactive} or augmented reality \cite{carmigniani2011augmented} for merging physical and virtual perception of a user. The former finds application in digital health whereas the latter is used in digital business, e.g., for virtual viewings of properties. From an abstract point of view, also data science falls within the category of human-computer interaction because a complex data analysis process involves many individual steps which may not be automatically connectable but requires human intervention, e.g., via an explanatory analysis \cite{em_jJuly010115}. 

A general question, also related to the example from the previous section about political decision making, is if one needs always a form of a human-computer interface for solving AI tasks in higher organizational or societal layers, as represented by health-related problems, the economy or cities, or if this is only needed in special cases? Classically, the ideal case seems to be an human-free AI system because this eliminates a potential bias of the subjectivity of humans. However, as the example about political decision making or the doctor-in-the-loop shows this is not clear for such problems.

\begin{figure}[!t]
\centering
\includegraphics[width=0.7\textwidth]{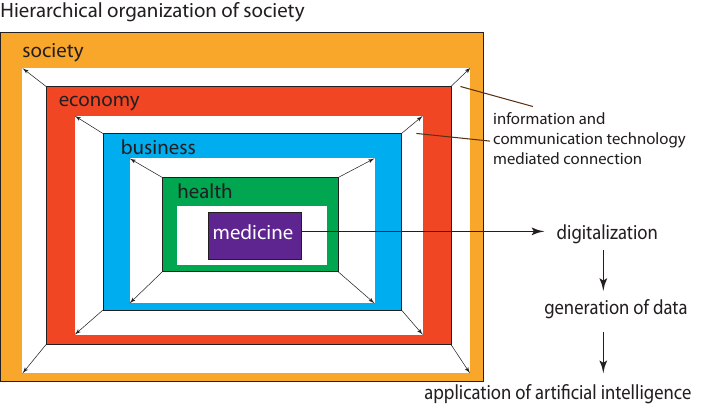}
\caption{A simplified view of the hierarchical organization of society. The digitalization of the shown fields progresses from the inside toward the outside. This leads eventually to a full penetration of society with artificial intelligence. }\label{fig.hierorg}
\end{figure}

\subsection{Data privacy and cybersecurity}

Another aspect which is also shared by the fields discussed in this paper is data privacy (information privacy) and cybersecurity. For instance, due to the increasing usage of connected technologies, e.g., IoT or CPS, Industry 4.0 systems are vulnerable to cyber attacks \cite{tuptuk2018security}. Similarly, patient data from hospitals or retirement homes contain sensitive personal information that need to be protected from third party usage. These problems are also well known for social media data, e.g., for using facebook or twitter \cite{isaak2018user,buccafurri2015comparing}.

A practical example where the usage of personal data by a third party may lead to unwanted consequences is the insurance or financial industry. For instance, private data could be used for the evaluation of an insurance premium or its coverage. Similarly, banks may base their decision to provide credits on similar data about and individuals. This problem is certain to become more severe the more the single layers of our society are interconnected with each other, see Fig. \ref{fig.hierorg}, in a way that the corresponding data from these layers become available.

A related issue to data privacy is artificial intelligence governance \cite{winfield2018ethical}. AI governance is concerned with ethical standards, safety, transparency and public fear. This addresses needs of individual users but also regulation bodies. Again there are cross links to our political decision making example above, especially relating the last point. For companies, on the other hand, there is the concern that AI governance might be too strict and rigid preventing the implementation of  viable business ideas. Overall, this will be a balancing act to find the right regulations satisfying the needs of all parties.

\subsection{From big data and cloud computing toward advanced analytics}

It is important to realize that despite the fact that data are the driving force (or fuel) of general AI methods they are not sufficient but necessary. This insight triggered the big data era where essentially all fields started to store all sorts of sector-specific data and cloud storage became popular. The next step that built upon big data was cloud computing because one needs not only data but also the capability to process these efficiently. Unfortunately, neither of these two steps generates value but they provide only the potential for deriving value thereof. In large, this can only be accomplished by the application of AI and machine learning methods. Hence, we need to move away from data storage and data processing as a main focus toward advanced analytics \cite{franks2012taming,halper2017advanced}. This includes visualization and dimension reduction methods for an explanatory data analysis \cite{ma2013review,tukey_1977} as well as unsupervised (clustering or hypothesis testing \cite{baldi2012autoencoders,em_jmake1030054}) and supervised learning (classification or regression \cite{bishop_2006,hastie_2009,em_jmake1010021}) methods. Furthermore, there are some modern learning paradigms that deserve more attention especially in combination with deep learning architectures \cite{kingma2014semi,li2017deep,weiss2016survey,han2019novel}: 
\begin{itemize}
\item semi-supervised learning
\item reinforcement learning
\item transfer learning
\item adversarial learning 
\end{itemize}

In addition, technology-mediated forms of AI are needed for optimizing the underlying technologies for digital health and digital economy, e.g., IoT, CPS or M2M.

On a statistical note, we would like to remark that, in general, the characteristics of a data set has a strong influence on the performance of any model. For this reason, also methods for investigating the statistical robustness of methods are of importance. This includes also resampling methods like cross validation (CV). In our discussion above we called such a field {\em statistical artificial intelligence}.

\subsection{From digital economy to digital society}

We want to finish our discussion with an outlook on the potential end-point of all these developments. What does this mean? Given the above explored information and communication technologies that find already application from digital health to digital economy, a natural question arising, is if it is possible to foresee where all this might lead us?

In our discussion about digital economy, we saw that digital economy is characterized by different sectors, of which Industry 4.0 is one of them. We think that this is a good way to indicate the continuous transformation or evolution of fields and would also suggest to utilize such an enumeration for digital economy itself, e.g., in the form of digital economy 4.0 \cite{skobelev2017way}. Thinking ahead, one may wonder what is the potential end-point of such a developmental process. Let's call this end-point digital economy X whereas 'X' corresponds to an unknown number. We don't know what the value of 'X' might be but it seems natural to assume that the limiting state or end-point of digital economy will provide a connected network of the individual layers manifesting society mediated by information and communication technologies (see Fig. \ref{fig.hierorg}). Hence, the resulting state has no borders between economy, business and society. That means every part of our home, work, education and recreation will be fully penetrated by AI and society will be also part of a business sector (see Fig. \ref{fig.hierorg} for a simplified visualization of such a hierarchical society). We call this end-point resulting from this development {\em digital society}.

\begin{formal}
\begin{quote}
    In a society that is fully connected by information and communication technologies enabling the pervasiveness of AI the asymptotic limiting state of digital economy X is called {\em digital society}.
\end{quote}
\end{formal}

From this scenario it seems clear that such a development would be undesirable because it would eradicate any kind of privacy. Hence, from an ethical point of view the question emerges what is the largest number 'X' we would be willing to tolerate?

It is interesting to note that there are similarities of our argument to Granovetter's theory of the embeddedness of economic actions in society via social networks, which he called {\it new economic sociology} \cite{granovetter1985economic}. This, in retrospective very plausible argument, however, penetrates also businesses and firms alike and in a fully connected world by information and communication technologies leads to a digital society.  

In general, detailed predictions about the future developments involving AI in our society are difficult \cite{dufva2019grasping}. Or to say it with the words of Stephan Hawking: "The rise of powerful AI will be either the best or the worst thing ever to happen to humanity. We do not yet know which". From reading this, it may not be surprising to find advocates along this broad spectrum of possible scenarios \cite{makridakis2017forthcoming,brockman2020possible}. For instance, Kurzweil represents an optimistic view that AI, along with nanotechnology and genetics, will improve our lives for the better, whereas Joy assumes a pessimistic role who even sees humanity threatened by such technologies to the extend of being extinct \cite{joy2000future}. We do not want to miss mentioning that there are also less extremal views on the future of AI. Such advocates may be seen as pragmatists because they belief in a beneficial application of AI by at the same time maintaining control over all crucial aspects of safety and security \cite{makridakis2017forthcoming}. The only thing that seems to be clear at this moment is that most results achieved so far are largely overhyped and we are still (far) away of full pervasiveness of artificial intelligence \cite{marcus2019rebooting}.

The best, and possibly most pragmatic, way to go forward seems to conduct more interdisciplinary research and education to enlighten our way ahead and to hopefully avoid devastating end-points \cite{helbing2019societal,Lankshear_2008}.

\section{Conclusion}

We hope that our perspective on the development of digital medicine and digital economy toward digital society, leading to a pervasiveness of artificial intelligence in all layers of society, demonstrates the need for a concerted effort in this area.

Finally, a point that seems to be underdiscussed in the current literature is 'responsibility'. Specifically, should AI scientists be responsible for their inventions and consequences these possibly have on society \cite{herrlich2013responsibility}? This can effect 'optimists' and 'pessimists' alike because from both negative consequences can arise in either over- or underutilizing opportunities. 

In summary, artificial intelligence in combination with digitalization offers a multitude of avenues we could go that could change our lives in profound ways. Interestingly, all of these ways seem to be inclusive with respect to the different scientific fields and application domains because AI raises question in technology, business and ethics alike. Hence, whatever the future will look like it will be multidisciplinary.

\vspace{6pt} 



\authorcontributions{FES conceived the study, wrote the manuscript and approved the final version.}



\conflictsofinterest{The authors declare no conflict of interest. The founding sponsors had no role in the design of the study; in the collection, analyses, or interpretation of data; in the writing of the manuscript, and in the decision to publish the results.} 

\reftitle{References}






\end{document}